\documentclass[%
 article,
 amsmath,amssymb,
reprint,%
]{revtex4-1}

\usepackage[version=4]{mhchem}
\usepackage{graphicx}% Include figure files
\usepackage{dcolumn}% Align table columns on decimal point
\usepackage{bm}% bold math
\usepackage[separate-uncertainty = true,multi-part-units=single]{siunitx}
\usepackage[utf8]{inputenc}
\usepackage[T1]{fontenc}
\usepackage{mathptmx}
\usepackage{etoolbox}
\usepackage{steinmetz}
\usepackage{mathtools}
\usepackage{overpic}
\usepackage{pstricks}

\makeatletter
\def\@email#1#2{%
 \endgroup
 \patchcmd{\titleblock@produce}
  {\frontmatter@RRAPformat}
  {\frontmatter@RRAPformat{\produce@RRAP{*#1\href{mailto:#2}{#2}}}\frontmatter@RRAPformat}
  {}{}
}%
\makeatother
\begin{document}

\preprint{AIP/123-QED}

\title{Improving magnetic-field resilience of NbTiN planar resonators using a hard-mask fabrication technique}
% Force line breaks with \\
\author{A. Bahr}
\author{M. Boselli}%
\author{B. Huard}%
\author{A. Bienfait}%
\affiliation{Ecole Normale Supérieure de Lyon,  CNRS, Laboratoire de Physique, F-69342 Lyon, France}%
\email{audrey.bienfait@ens-lyon.fr}

\date{\today}

\begin{abstract}
High-quality factor microwave resonators operating in a magnetic field are a necessity for some quantum sensing applications and hybrid platforms. Losses in microwave superconducting resonators can have several origins, including microscopic defects, usually known as two-level-systems (TLS). Here, we characterize the magnetic field response of NbTiN resonators patterned on sapphire and observe clear absorption lines occurring at specific magnetic fields. We identify the spin systems responsible for these features, including a yet unreported spin with $g=1.85$ that we attribute to defects in the NbTiN thin film. We develop mitigation strategies involving namely an aluminum etch mask, resulting in maintaining quality factors above $Q>2 \times 10^5$ in the range \SIrange{0}{0.3}{T}.
\end{abstract}

\maketitle

Superconducting circuits are promising candidates for quantum information processing (QIP), as well as high-precision sensing. Recently, the interest in circuits able to operate under magnetic fields has increased to enable hybrid QIP platforms\cite{burkardSuperconductorSemiconductorHybridcircuit2020,clerkHybridQuantumSystems2020} or quantum sensing applications such as magnetic resonance detection\cite{sigillitoFastLowpowerManipulation2014,wangSingleElectronspinresonanceDetection2023}. These applications require developing high-quality factor superconducting resonators resilient to magnetic fields. Detrimental effects for microwave superconducting resonators have been largely studied \cite{mcraeMaterialsLossMeasurements2020}, with evidence that obtaining high-quality factors at zero magnetic field requires shielding from quasi-particles\cite{barendsMinimizingQuasiparticleGeneration2011a}, preventing vortex nucleation\cite{songMicrowaveResponseVortices2009}, using low-loss dielectric substrates\cite{wangSurfaceParticipationDielectric2015,dialBulkSurfaceLoss2016}, and mitigating microscopic defects\cite{mullerUnderstandingTwolevelsystemsAmorphous2019}. This last effect arises from so-called two-level-systems (TLS) which can be saturated to recover a higher quality factor at high power\cite{barendsNoiseNbTiNTa2009}. They can have different origins, including spurious spin systems\cite{mullerUnderstandingTwolevelsystemsAmorphous2019}. When the spin Larmor frequency is tuned on resonance by the application of a magnetic field, an increase in losses can be observed which allowed to identify these spin systems partly as hydrogen and oxygen adsorbates on sapphire substrates\cite{quintanaObservationClassicalQuantumCrossover2017,degraafDirectIdentificationDilute2017}. In addition to representing a source of losses when operating under magnetic field, these spins are also believed to be responsible for frequency noise as well as flux noise \cite{barendsNoiseNbTiNTa2009} limiting for example the sensitivity of SQUID magnetometers\cite{antonMagneticFluxNoise2013,sendelbachMagnetismSQUIDsMillikelvin2008a} or the coherence of flux qubits\cite{sankFluxNoiseProbed2012,quintanaObservationClassicalQuantumCrossover2017,kumarOriginReductionMagnetic2016a}. Here, we evidence similar field-dependent losses for superconducting resonators deposited on sapphire, and show that a hard-mask nanofabrication technique mitigates the contribution of some of these spurious spin systems.

\begin{figure}[htpb!]
\includegraphics{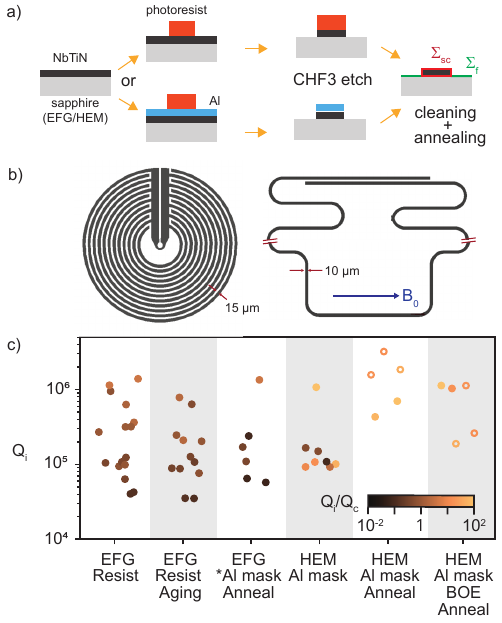}
\caption{\label{fig:fig1}(a) Fabrication steps for NbTiN superconducting resonators: 50-nm films sputtered on EFG or HEM grown sapphire, coated by a resist or Al mask, dry etched with \ce{CHF3} chemistry, and then cleaned and possibly annealed at \qty{300}{\degreeCelsius} under \ce{N2} atmosphere. (b) Resonator geometries used in this work, nicknamed circle (left) and microwire (right). The microwire drawing is cut along the red lines for clarity. The blue arrow $B_0$ shows the applied in-plane magnetic field. (c) Measured intrinsic quality factors at zero field, depending on the fabrication technique, filled (open) markers indicate circle (microwire) design, their colors indicate their $Q_i/Q_c$ ratio as given by the color bar in inset. X-coordinate scattering is only for clarity. *Al mask indicates the wafer was covered by photoresist during dicing, but that an Al mask was used during etching.}
\end{figure}
TLS defects can be either electric dipoles or magnetic dipoles whose interaction strength with the resonator field depends on their location   \cite{mullerUnderstandingTwolevelsystemsAmorphous2019}, and range from weak to strong coupling\cite{lisenfeldDecoherenceSpectroscopyIndividual2016}. These defects can be located in the substrate or at the surfaces of the substrate and superconducting electrodes. Electric-TLS can be for instance trapped charges, dangling bonds or tunneling atoms located either on the substrate or in an oxide layer capping the superconductor. Many works have now demonstrated that surface dielectric losses are far more detrimental than dielectric bulk losses\cite{wangSurfaceParticipationDielectric2015, woodsDeterminingInterfaceDielectric2019}, and shown that the impact of these electric-TLS can be minimized by improving materials and fabrication techniques as well as optimizing the design to reduce their contribution \cite{mullerUnderstandingTwolevelsystemsAmorphous2019, mcraeMaterialsLossMeasurements2020, murrayMaterialMattersSuperconducting2021}. 

Magnetic-TLS arise from nuclear and electronic spins lying at the surfaces or in the bulk of the substrate, such as spin donors for silicon\cite{ledererSubstrateLossMechanisms2003}, or metal contaminants in sapphire\cite{farrUltrasensitiveMicrowaveSpectroscopy2013}.  Their contribution to the resonator intrinsic loss and flux noise is maximum when their transitions are brought to resonance by the application of a magnetic field. Several superconductors, such as \ce{Nb}\cite{kwonMagneticFieldDependent2018}, \ce{NbN}\cite{degraafDirectIdentificationDilute2017,yuMagneticFieldResilient2021}, \ce{NbTi}\cite{russoFabricationCharacterizationNbTi2023}, \ce{NbTiN}\cite{samkharadzeHighKineticInductanceSuperconductingNanowire2016},  GrAl\cite{borisovSuperconductingGranularAluminum2020} and \ce{YCBO}\cite{ghirriYBa2Cu3O7MicrowaveResonators2015a}, have proven able to yield high-quality factor resonators ($Q>10^5$) at moderate ($<1$~T) and large magnetic fields (\qty{6}{T}). Yet, in most works reporting resonators resilient to magnetic fields, a drop in their intrinsic quality factors appears at specific magnetic fields. Using resonators of various frequencies fabricated in \ce{NbTiN} on Si, \ce{NbN} on \ce{SiO2}, and granular Al on sapphire, previous works\cite{samkharadzeHighKineticInductanceSuperconductingNanowire2016,yuMagneticFieldResilient2021,borisovSuperconductingGranularAluminum2020} observed losses corresponding to a near perfect spin $1/2$ bath with a g-value $g= \hbar\gamma_{\mathrm{eff}}/\mu_{\mathrm{B}} = 2.0 \pm 0.3$, where $\gamma_{\mathrm{eff}}=\partial \omega_{\mathrm{s}}/\partial B_0$ is the  effective gyromagnetic ratio governing the measured spin frequency $\omega_{s}$ dependence on the applied magnetic field $B_0$. Besides, Kroll and coworkers\cite{krollMagneticFieldResilientSuperconductingCoplanarWaveguide2019} also observed a similar effect for \ce{NbTiN} on sapphire, but at a value slightly lower than 2 (about g = 1.9). For \ce{NbN} on sapphire\cite{degraafDirectIdentificationDilute2017},  a spin 1/2 absorption line $\omega_s(B_0)$ was observed, along with satellite lines corresponding to hydrogen physiosorbed on the surface, which creates a spin system with a signature hydrogen hyperfine splitting of $A/(2\pi) = $\qty{1.42}{GHz}. These hydrogen atoms absorbed on the surface are also thought to contribute significantly to flux noise in \cite{quintanaObservationClassicalQuantumCrossover2017} in superconducting circuits. Additionally, contribution from \ce{O2} adsorbates and superoxide systems embedded into the surface of the sapphire lattice was reported\cite{degraafDirectIdentificationDilute2017}, further confirmed by high-field EPR studies \cite{unNatureDecoherenceQuantum2022} and DFT calculations \cite{leeIdentificationLocalSources2014}. They show that spin systems respectively create a broad absorption at $g \approx 2.04$ and constant losses occurring for $g \in [1.5,4]$. Here, we show that using an aluminum mask instead of a photoresist mask when patterning \ce{NbTiN} thin films prevents the appearance of most of these TLS, except for a signal we attribute to defects within the NbTiN films.
\begin{figure}
\includegraphics{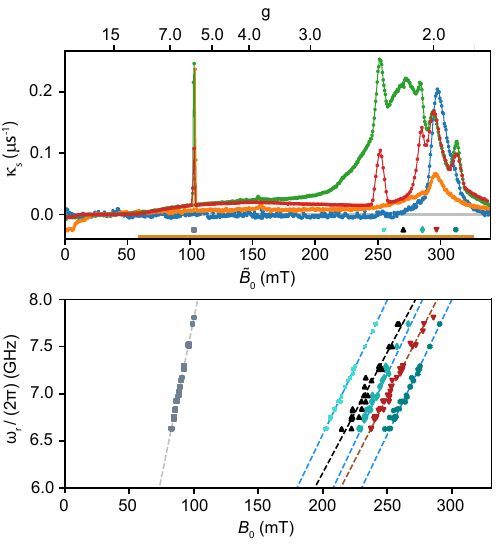}
\caption{\label{fig:fig2p2} (a)  Magnetic-field induced dissipation rate of circle resonators for different fabrication runs as a function of magnetic field, error bar per point is less than \qty{5}{\per\milli\second}. Red, NbTiN on EFG substrate etched through a resist mask ("EFG Resist"); green, same sample with 2 days aging ("EFG Resist Aging"); orange, change to an Al mask with annealing ("EFG Al mask Anneal"); blue, change to HEM substrate, no resist coating during dicing ("HEM Al mask Anneal"). For direct comparison, the magnetic field is rescaled (see main text). (b) Frequency-diagram of all absorption peaks:  each marker y-coordinate is given by the frequency of a resonator, and its x-coordinate corresponds to the occurrence field of a given feature.  Blue dashed lines model an electronic spin $1/2$ coupled to a nuclear spin $1/2$ with a hydrogen hyperfine coupling, gray, black and red lines are linear fits corresponding respectively to $\Delta_0/(2\pi)= 1.07, 1.04, 0.28$~GHz and $g = 4.66, 1.77, 1.85$.}
\end{figure}

We pattern \ce{NbTiN} resonators using a dry etching process (Fig. 1a) by starting from a 50-nm-thick \ce{NbTiN} film sputtered on a 2" sapphire wafer. The wafer is either an edge-defined-grown wafer (EFG, University Wafers), or grown by a heat-exchanger method (HEM, Crystec). The wafer is then diced in rectangular dies of \qty{5}{\mm} $\times$ \qty{7}{mm}, protected using either photoresist (EFG wafers), or a 50-nm-thick Al layer covered in photoresist (HEM wafers). For each die, we pattern using optical lithography either a photoresist mask (Shipley S1813) or an aluminum mask. The latter is realized by safekeeping the Al layer deposited pre-dicing, or by depositing 50-nm of aluminum after cleaning the photoresist off the die post-dicing. The pattern from the photoresist is transferred to the Al layer by an acid wet-etching (Transene Type A) and the photoresist is cleaned off. The \ce{NbTiN} layer is then etched using an ICP-RIE tool (Sentech) using \ce{CHF3}/\ce{Ar}/\ce{N2}/\ce{O2} atmosphere (30/100/10/3 sccm). If an Al mask is present, it is then wet etched with the same etch as above. A final cleaning is then performed using N-Methyl-2-pyrrolidone followed by isopropanol rinsing. On some devices, to remove the identified contribution of the physiosorbed hydrogen, we performed an annealing at \qty{300}{\degreeCelsius} or \qty{600}{\degreeCelsius} in \ce{N2} atmosphere. On one die, we performed a buffered oxide etch of 20 min at room-temperature to remove the oxides formed on top of the \ce{NbTiN} layer before performing an annealing in \ce{Ar} atmosphere. In total, we prepared 9 dies, with various preparation conditions summarized in the supplementary material. Out of these 2 dies, two were subjected to open air aging and were measured twice.

We have used two types of resonator geometry: low impedance resonators ($Z_{\mathrm{c}} \approx 20-40~\Omega$) targeted for electron spin resonance (ESR) sensing (circle), and higher impedance ($Z_{\mathrm{c}} \approx 200 ~\Omega$) resonators targeted for resilience to magnetic field (microwires) see Fig 1b. The first geometry comprises a circular interdigitated capacitor of pitch \qty{20}{\um} surrounding an inductive loop or a short microwire. The second is an inductive ribbon of width \qty{10}{\um} whose ends are capacitively coupled. The resonators are capacitively or inductively coupled to a coplanar waveguide (CPW). The measurement is done either  in hanger geometry with the CPW running through the chip, or in reflection with the CPW terminated on chip. The width of the ground plane surrounding the CPW is reduced to \qty{70}{\um}  so that all resonators are free floating and not encased in a superconducting ground plane. This is done to minimize the nucleation of magnetic vortices nearby the resonators \cite{songMicrowaveResponseVortices2009} and prevent vortices induced losses when applying a dc magnetic field.

On each die, 6 to 8 resonators were patterned. Using well-established fitting procedures \cite{megrantPlanarSuperconductingResonators2012,probstEfficientRobustAnalysis2015,riegerFanoInterferenceMicrowave2023}, we extract for each resonance its frequency $\omega_{\mathrm{r}}/2\pi$, and its intrinsic and coupling quality factors $Q_i$ and $Q_c$ at low intra-resonator photon number  ($\bar{n}\sim 1-10$). The intrinsic quality factors $Q_i^0$ obtained at zero magnetic-field for the different fabrication techniques are shown in Fig.~1c and range from $10^4$ to $10^6$. All fabrication techniques exhibit at least a few resonators with $Q_i$ exceeding $5\times 10^5$. All devices fabricated using HEM wafers, and whose NbTiN layer was not in contact with resist, obtain $Q_i$ consistently larger than $10^5$ at zero magnetic fields. With annealing, the intrinsic quality factors improve slightly.

We next study the behavior of each resonator under magnetic field. To isolate the dependence of the resonator intrinsic decay rate $\kappa_{\mathrm{s}}$ on the magnetic field, we subtract the loss rate $\omega_{\mathrm{r}}/Q_i^0$ observed at zero-field and the additional loss rate $\omega_{\mathrm{r}}/Q_i^{wb}$ occurring when the Al wire bonds connecting the die to the sample holder transit from superconducting to normal state at $B_0 \sim $\qty{10}{mT}.  We plot $\kappa_{\mathrm{s}}(\tilde{B}_0)$ in Fig. 2a for resonators made with different fabrication techniques, using a re-scaled x-axis $\tilde{B}_0=B_0~ \tilde{\omega}_{\mathrm{r}}/\omega_{r}$ with $\tilde{\omega}_{\mathrm{r}}/2\pi=$\qty{8}{GHz} for easier comparison.  Focusing first on a NbTiN resonator patterned on an EFG wafer using a photoresist mask (green), we observe 4 different peaks, as well as the onset of a large plateau at $\tilde{B}_0 = $\qty{50}{mT}. To identify these features, we build their frequency-field diagram using all the resonators we have measured (see Fig.~2b). We find that the absorption peak frequency of each feature depends linearly on the magnetic field, allowing us to extract a zero-field splitting $\Delta_0$ and an effective $g$ value for each feature: $\omega_{\mathrm{r}} = \Delta_0 + g \frac{ \mu_B}{\hbar} B_0$.

The first peak lying at $\tilde{B}_0 =$\qty{100}{mT} is very sharp compared to the others, and corresponds to a $g$-value of $4.7$ (gray, Fig.~2b). We attribute it to a response from the sapphire substrate. Indeed, while this effective $g$-value does not correspond to typical contaminants (\ce{Fe},\ce{Cr}), less typical impurities (\ce{V},\ce{Mo},\ce{Mn},\ce{Ti},\ce{Gd}) or known defects in sapphire which are magnetically active (\ce{V} and \ce{F2} centers), this line has already been observed without being identified in other studies using HEMEX sapphire \cite{farrUltrasensitiveMicrowaveSpectroscopy2013} and Verneuil grown sapphire \cite{bletskanDeterminingResidualImpurities2008}. We thus note that it cannot be tied to a particular growth technique. We nevetheless observe its disappearance when switching from EFG to HEM-grown sapphire. Note that we also observe a faint line at $g = 3.3$  that also disappears and that we cannot identify.

Next, we turn to the spin 1/2 line surrounded by an upper and lower satellite peaks (blue markers in Fig.~2b). These three lines have been observed in  Ref\cite{degraafDirectIdentificationDilute2017} and have been attributed to water adsorption on the sapphire surface resulting in imperfect hydroxylation of the surface and the creation of radicals and unpaired hydrogens. The radicals create the central spin 1/2 line ($g=2$), and the coupling of a large part of these radicals to atomic hydrogen gives rise to a hyperfine splitting of strength $A/(2\pi) = 1.45$ GHz leading to the appearance of the satellite peaks. These features can be removed by desorbing the sapphire surface, which was successfully done in previous work\cite{unNatureDecoherenceQuantum2022} by annealing at \qty{300}{\degreeCelsius} under vacuum.  We observe a similar disappearance using an annealing at the same temperature under both \ce{N2} or \ce{Ar} atmosphere in our measurements. Using the relative amplitude of the two satellite peaks, we can infer the polarization of the probed spin ensemble and deduce a spin temperature of \qty{80}{\milli\kelvin}. This value is reasonable considering the limited attenuation we have placed on the input probe line (see supplementary material).

We now move to two correlated features linked to the adsorption of \ce{O2} and \ce{H2O} on the sample surface: the peak at  $\tilde{B}_0 =$\qty{270}{\milli\tesla} and the plateau appearing at  $\tilde{B}_0 =$\qty{50}{\milli\tesla}.  Imperfect hydroxylation and defect sites in the sapphire located near the surface leads to the production of superoxides (\ce{O2-} and \ce{HO2-}) contributing to the broad peak at $g = 1.77$ (black markers in Fig.~2b)\cite{unNatureDecoherenceQuantum2022,degraafDirectIdentificationDilute2017}. For \ce{O2} molecules which are not reduced by interaction with the surface (either the sapphire or the NbTiN surface), they are expected to respond to ESR probing accordingly to their triplet state nature i.e., as a broad plateau resonance extending from $g = 4.0$ to low values. We see both features in our experiment, with the superoxide peak appearing more strongly for aged samples which were left in open air for 3 days.

\begin{figure}
\includegraphics{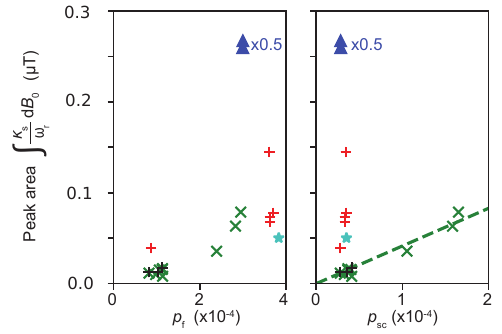} % same as 9 but axis for pr switched

\caption{Integrated losses for the $g=1.85$ peak depending on the participation ratio of the free sapphire surface (left) and of the NbTiN surface (right). Markers indicate various fabrication conditions: etch through an aluminum mask without post-annealing (green crosses), aluminum mask with post-annealing at \qty{300}{\degreeCelsius} when the NbTiN was covered with resist during dicing (black crosses) or was never in contact with resist (red crosses), etch through aluminum mask with post-annealing at \qty{600}{\degreeCelsius} (blue triangles, $\times0.5$ for clarity), etch through an aluminum mask with argon post-annealing at \qty{300}{\degreeCelsius} (cyan star). The green dash line is a linear fit using the black and green crosses, enabling to find the spin concentration giving rise to the  $g=1.85$ peak. \label{fig:fig2p2}}
\end{figure}

The last feature lies at $g=1.85$, with an extracted zero-field splitting $\Delta_0 = 0.28 \pm 0.12$ GHz (red markers in Fig.~2b). The absorption feature has an asymmetric shape, and has not been previously reported, though it may have been identified as a spin 1/2 line in a similar study of NbTiN resonators on sapphire \cite{krollMagneticFieldResilientSuperconductingCoplanarWaveguide2019}. To identify its origin, we calculated for each resonator geometry the participation ratio $p_i$ of each surface $\Sigma_i$  to the total mode energy using finite-elements microwave simulations (see supplementary material) \cite{wangSurfaceParticipationDielectric2015}. Namely, we wish to distinguish between two hypotheses: whether this spin system is located on the sapphire free surface $\Sigma_f$, or on the superconductor surface $\Sigma_{\mathrm{sc}}$ (see Fig.~1a). Using a multipeak fitting routine, we have calculated for each resonator the integrated absorption loss $\int \kappa_{\mathrm{s}}/\omega_0 ~dB_0$. In Fig.~3, we plot this peak area versus $p_{\mathrm{f}}$ and $p_{\mathrm{sc}}$ for various fabrication protocols. Concentrating on the fabrication without annealing (green markers) for both a resist or an aluminum mask, we see that the peak area is proportional to $p_{\mathrm{sc}}$, while its dependence on $p_{\mathrm{f}}$ is non-linear. We thus conclude that the defect is more likely to emanate from the NbTiN surface. From this linear fit, we can extract that the surface spin concentration is \qty{3.1(2)e12}{\per\centi\meter\squared} (see supplementary material). Besides, the absorption does not depend on the in-plane orientation of the magnetic field so that the spins are either isotropic, or randomly orientated. We can explore a few hypotheses on the origin of these spins.  The fluorine etch could leave paramagnetic residues on the edge of the superconducting electrodes\cite{fonashOverviewDryEtching1990,surajitkumarhazraGetteringCF4ArPlasmaTreated2009}. For instance, \ce{TiF3} has a paramagnetic response, yet its g-value is anisotropic and would give rise to a broad line extending from 2 to 1.85\cite{voreTitaniumDifluorideTitanium2002}, which does not agree with our measurements. Some electronic states of niobium and titanium exhibit a paramagnetic response, for instance  electronic states \ce{Nb2+} and \ce{Nb4+} have a paramagnetic response with $g$-values around $1.85$\cite{johnsonElectronSpinResonance1973, ziolekCharacterizationTechniquesEmployed2003}, while \ce{Ti3+} is anisotropic, with principal $g$ -values lying between 1.9 and 2. Yet, we note that the feature does not appear in \ce{Nb} or \ce{NbN} resonators \cite{kwonMagneticFieldDependent2018,degraafDirectIdentificationDilute2017}, so a contribution from \ce{Nb} donor states seems unlikely, or deeply linked to the presence of \ce{Ti} in the film. 

These features contribute to a background signal strongly dependent on the magnetic field, which complicates the identification of an ESR signal, and that also limits the overall achievable quality factor. We now review the impact of different fabrication processes on the different features to elaborate a mitigation strategy. We plot on Fig.~4 the peak height $\kappa_{\mathrm{s}}^f$ for each feature, depending on the fabrication protocol. The contribution for each feature was isolated using a multi-peak fit.  (1) We note that similarly to measurements on NbN resonators on sapphire, annealing removes the hydrogen peaks (blue) as well as the central \ce{OH-} peak, independently of the annealing gas.  (2) Avoiding contact between photoresist and NbTiN suppresses  the plateau resulting from the contribution of triplet \ce{O2}: indeed for HEM wafers where only Al was in contact with NbTiN, the plateau is absent, but for EFG wafers, even when an Al mask is used, contact with the protective photoresist layer during dicing is enough for the plateau to be observed. (3) When cleaning the wafer with a long buffered oxide etch after removing the mask, the plateau reappears, with a reduced height. It also gives rise to a small $g=2$ feature which is not removed by annealing. Putting aside the \ce{NbTiN} feature, it thus looks like the optimal fabrication technique among those explored here is etching through an aluminum mask followed by an annealing.

\begin{figure}
\centering
\includegraphics{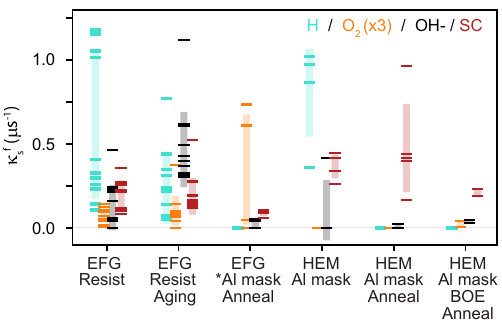}

\caption{Feature peak height $\kappa_{\mathrm{s}}^{\mathrm{f}}$ extracted from a multipeak fit for the following features: lowest hydrogen hyperfine peak (blue), \ce{O2} plateau (orange, $\times 3$, for visibility), \ce{OH-} radicals (black), and the unidentified feature related to the superconducting film (red) depending on each fabrication protocol. Markers indicate measured values, boxes show the standard deviation around the mean. *Al mask indicates the wafer was covered by photoresist during dicing, but that an Al mask was used during etching.\label{fig:fig4} 
}
\end{figure}

The \ce{NbTiN} feature at $g=1.85$ however shows a strong increase when the sample is annealed : for samples whose NbTiN was never covered by resist, we observe a factor $6$ increase for a \qty{300}{\degreeCelsius} annealing under \ce{N2} or \ce{Ar} atmosphere (red), and a factor $45$ increase  for \qty{600}{\degreeCelsius} annealing under \ce{N2} (see Fig.~3). Note that we observe that when heating NbTiN under \ce{N2} atmosphere at higher temperature, the film swells and gets enriched in nitrogen, moving away from a balanced two atoms face cubic-centered lattice with half of the lattice occupied by nitrogen atoms, and the other occupied by \ce{Nb} and \ce{Ti} atoms \cite{zhangSuperconductingPropertiesChemical2015}.  It could thus be that enrichment of the films in nitrogen and/or a depletion of the film in \ce{Ti} creates donor niobium sites that give rise to the feature. Remarkably, the sample which was covered by resist for dicing, but etched with an aluminum mask does not show the same increase so that we postulate that an oxide passivation layer is sufficient to counteract the effect of annealing on the \ce{NbTiN} feature.

Nevertheless, avoiding putting photoresist into contact with \ce{NbTiN} for dicing and etching through an aluminum mask followed by an annealing results in the best combination to remove all magnetic field dependent features. It also ensures that we reach reproducible quality factors above $Q>10^5$ at zero magnetic field, in particular for our low-impedance geometry resonator. Considering all sources of losses, including the contribution from aluminum wire bonds ($\sim$\qty{3e5}) and of the \ce{NbTiN} feature ($\sim$\qty{2e5}), we can maintain a total quality factor above $Q > 10^5$ in the \SIrange{0}{0.3}{\tesla} range, an important step to develop an ESR spectrometer based on superconducting microwave resonators. Straightforward improvements to our setup could help reduce the length of the wire bonds so that their contribution is minimized ($>$\qty{2e6}). The behavior of the \ce{NbTiN} feature also makes clear that developing a capping layer that would act as a barrier during the annealing without contributing to the oxygen features would help us reach higher quality factors. Broader perspectives would be to probe the origin of the \ce{NbTiN} feature by varying the stoichiometry of the \ce{NbTi} sputtering target or comparing to \ce{TiN} resonators.

\begin{acknowledgments}
This work was supported by the LABEX iMUST (ANR-10-LABX-0064) and the project IDEXLYON of Université de Lyon, within the program "Investissements d'Avenir" (ANR-11-IDEX-0007 and ANR-16-IDEX-0005) operated by the French National Research Agency (ANR) and by the European Union (ERC, INDIGO, 101039953). We acknowledge IARPA and Lincoln Labs for providing a Josephson Traveling-Wave Parametric Amplifier. The NbTiN films were etched using the ICP-RIE tool of the Nanolyon platform. We thank the cleanroom of the SPEC at CEA-Saclay and Pief Orfila for sputtering NbTiN on our wafers.

\end{acknowledgments}

\section*{Data Availability Statement}
The data that support the findings of this study are available from the corresponding author upon reasonable request.

% The data that support the findings of this study are openly available in zenodoe at http://doi.org, reference number XXX.

\section{Appendixes}

\subsection{Samples and measurement setup}
The list of samples as well as their precise fabrication process is given in Table~\ref{tab:sample}.

\begin{table*}[ht]
    \centering
\begin{tabular}{|c|c|c|c|c|c|c|c|c|}
     \hline
     Sample & Substrate & Dicing & Cleaning & Mask & Etch & Cleaning & Annealing & Probe geometries \\ \hline
     1*& EFG & Resist & NMP & Resist & \ce{CHF3} & O2 plasma \& NMP & No & Reflection \\
     2*& EFG & Resist & NMP & Resist & \ce{CHF3} & O2 Plasma \& NMP & No & Reflection \\
     3& EFG & Resist & NMP & Resist & \ce{CHF3} & NMP & No & Reflection \\
     4& EFG & Resist & NMP & Al & \ce{CHF3} & Al etch& \qty{300}{\degreeCelsius} \ce{N2}  & Reflection \\
     5& HEM & Al  + Resist & NMP & Al & \ce{CHF3} &  Al etch& \qty{300}{\degreeCelsius} \ce{N2}  & Hanger \\
     6& HEM & Al  + Resist & NMP & Al & \ce{CHF3} & Al etch& No & Reflection \\
     7& HEM & Al  + Resist & NMP & Al & \ce{CHF3} &  Al etch& No & Reflection \\
     8& HEM & Al  + Resist & NMP & Al & \ce{CHF3} & Al etch& \qty{300}{\degreeCelsius} \ce{Ar}  & Hanger \\
     9& EFG & Resist& NMP & Al & \ce{CHF3} & Al etch& \qty{600}{\degreeCelsius} \ce{N2} & Hanger \\
     \hline
\end{tabular}
    \caption{List of measured samples and their fabrication process. Samples 1 and 2 were measured in a second run after a few days ageing.}
    \label{tab:sample}
\end{table*}
After fabrication, the samples are wirebonded to a OFHC copper sample holder and inserted into a vector magnet anchored to the base plate of a dilution refrigerator operating at \qty{12}{mK} for measurement. The microwave setup is shown on Fig.~5. The magnet is placed inside a cryoperm shield and calibrated using BDPA samples. The incoming signal is attenuated with a total of \qty{43}{\dB}. We include an Eccosorb filter in-front of the circulator before the device under test (DUT) to reduce quasi-particles caused by infrared radiation. The input power is chosen such that the intra-resonator photon number is of order 1. The outgoing signal is amplified by a Josephson traveling-wave parametric amplifier followed by a high-electron mobility transistor amplifier. The measurements are carried out using a vector network analyzer (VNA).
\begin{figure}[t]
	\centering
	\includegraphics[width=0.3\textwidth]{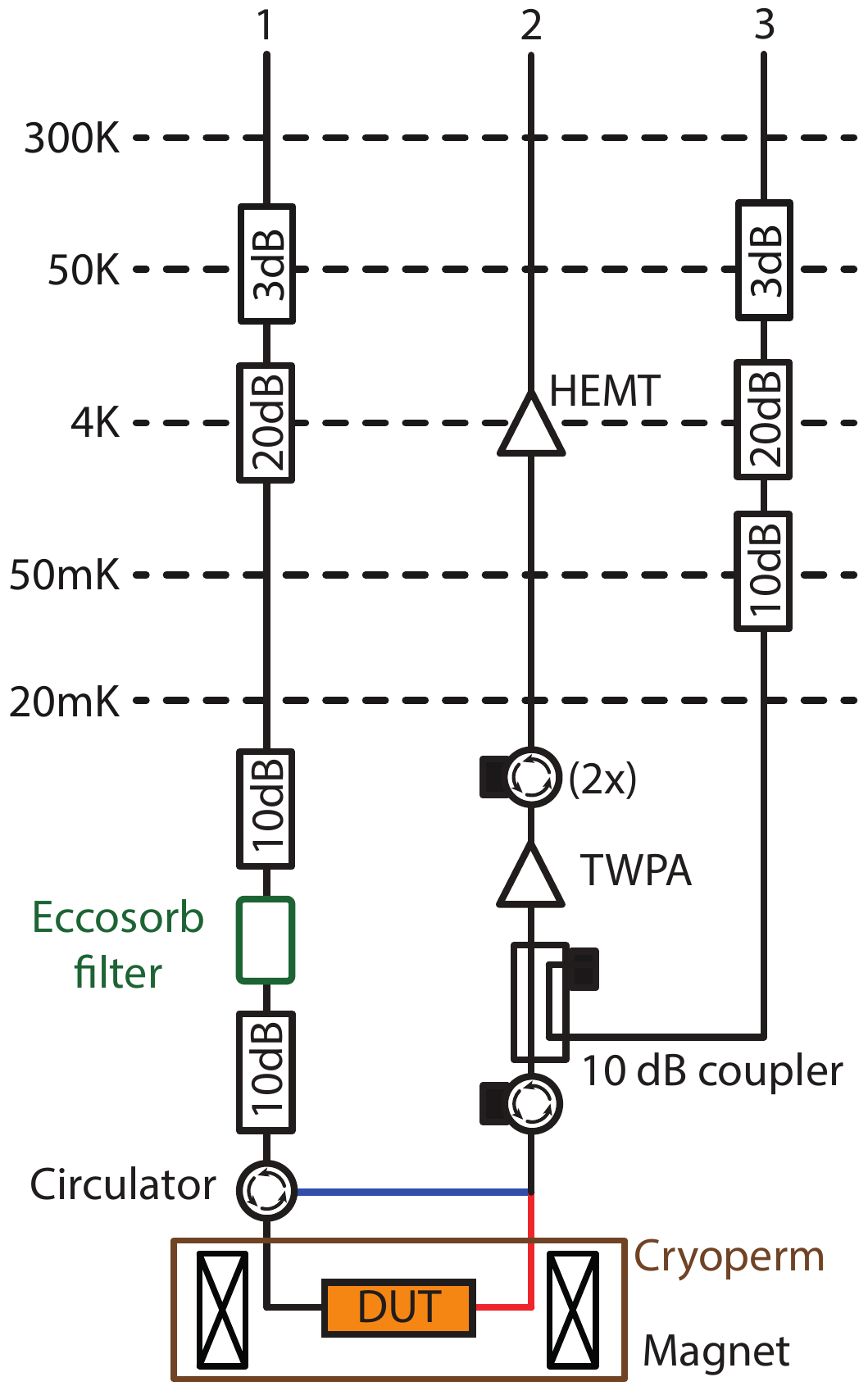}
	\caption{Microwave setup for transmission (red) and reflection measurements (blue).}
	\label{fig:FridgeSetup}
\end{figure}%

\subsection{Resonator fitting}
The resonators on the various samples are probed either in reflection or in a transmission hanger geometry. To extract the internal quality factor from the VNA measurements for these two coupling schemes, we model the reflection ($R_R$) and hanger-mode ($R_H$) response as \cite{probstEfficientRobustAnalysis2015,riegerFanoInterferenceMicrowave2023}:
\begin{equation}
    \begin{split}
        R_R &= \left[1-\frac{2Q_l/Q_c}{1+2i  Q_l(\omega/\omega_{\mathrm{r}}-1)}\right]\\
        R_H &= \left[ 1-\frac{\left(Q_l/Q_c\right) e^{i\phi_0}}{1+2iQ_l(\omega/\omega_{\mathrm{r}}-1)}\right]
    \end{split}
 \label{eqn:scattering_parameters_VNA}
\end{equation}

The complex phase $\phi_0$ accounts for impedance mismatches in the coupling between resonator and measurement lines. We define $Q _l = \left[1/Q_i + 1/Q_c \right]^{-1}$ as the loaded loss of the resonator. The measured response of the resonator $S_m$ at the VNA is actually convoluted with the response of the measurement lines. This response can be  described as an overall complex scaling factor $Ae^{i\alpha}$  and an electrical delay $\tau$ giving a frequency-dependent phase $e^{i\omega\tau}$. To recover only the resonator response, we renormalize the data using  $\tilde{S}_m = S_m(\omega) e^{-i \omega t } /S_m(\infty)$ so that $\tilde{S}_m(\omega)$ lies on a circle in the complex plane.
We then estimate the best circle fitting the data in the complex plane, obtaining its radius $R$ and center \mbox{$z_c = x_c + i y_c$}. 

We perform the affine transformation \mbox{$ S_{21}^{\{ R,H\} _C} = S_{21}^{\{ R,H\} _N}-z_c$} and compute the phase of the transformed quantity, obtaining:
\begin{equation}
    \begin{split}
\phi_R &= 2 \arctan\left[2(\omega/\omega_{\mathrm{r}} - 1)Q_l\right]\\
\phi_H &= 2 \arctan\left[2(\omega/\omega_{\mathrm{r}} - 1)Q_l\right]-\phi_0
    \end{split}
\end{equation}
which can be fitted to extract the loaded quality factor and the resonant frequency of the resonator. Using the relations between ($Q_i$,$Q_c$) and ($R$,$Q_l$)
\begin{equation}
    \begin{split}
    Q_c^R = Q_l ^R / R \hspace{1 cm}& 1/Q_i^R = 1/Q_l^R - 1/Q_c^R\\
    Q_c^H = Q_l ^H / 2R \hspace{1 cm}& 1/Q_i^H = 1/Q_l^H - 1/Q_c^H
    \end{split}
\end{equation}
it is possible to estimate the coupling and internal quality factors. 
As the determination of the quality factors involves three steps (re-normalization, circle fitting, and phase fitting), each step can contribute to the uncertainty. In practice, the errors due to the circle fitting, and in particular on the circle radius largely offset the other two. The Qc value is fixed to the average fitted value for all magnetic fields thus reducing its uncertainty. For the intrinsic quality factor, as $1/Q_i = (1-r R)/(r R Q_c)$  we estimate $\Delta(1/Q_i) = \Delta(R)/(r R^2 Q_c)$ where $r=2$ in hanger geometry and $r=1$ in reflection. This analysis does not take into account Fano interferences or direct-direct port leakage which can introduce a systematic error in the determination. However, as we are only looking at extra-losses created into resonators of similar geometries, this systematic error does not particularly affect our study.

\subsection{Modelling of surface spin losses for a superconducting resonator}

In this section, we model losses created by spins lying either on the metallic area of a planar superconducting resonator, or on the nearby uncovered substrate surface.
\subsubsection{Spin induced losses}
Each family of spins can be described by its Hamiltonian $H_s(\vec{B})$ which can be diagonalized to extract the spin family eigenstates and eigenvalues. When a static magnetic field $\vec{B}_0$ is applied, a specific transition between states $\|1\rangle$ and $|2\rangle$ occurring at frequency $\omega_s(B_0)$ can be brought to resonance. Each spin interacts with the resonator through a Jaynes-Cumming Hamiltonian. The interaction strength $g_j$ of a single spin is given by $\hbar g = |\langle 1 | H_s(\vec{\delta B}(\vec{r}))|2\rangle|$, where $\vec{\delta B}_{ac}(\vec{r})$ are the vacuum fluctuations of the resonator electromagnetic mode at the spin position $\vec{r}$. We consider that the spin ensemble has an inhomogenous distribution in frequency of linewidth $\Gamma$ while each spin has a decoherence rate $\Gamma_2$. We also introduce the collective coupling constant as $g_{ens} = \sum_j g_j^2$.
In the limit of small cooperativity $4g_{ens}^2/(\kappa\Gamma)<<1$,  assuming the spin frequency distribution is Lorentzian, the response $R(\omega)$ for the resonator is \cite{dinizStronglyCouplingCavity2011a}:
\begin{equation}
    R(\omega) = 1- \frac{r \sqrt{\kappa_c}}{\kappa + \kappa_{\mathrm{s}} - 2i(\omega-\omega_{\mathrm{r}}) + 2 i \delta_s}\label{resoResp}
\end{equation}
where $\kappa_{\mathrm{s}} = 4g^2_{ens} \Gamma/((\omega_s-\omega_{\mathrm{r}})^2 + \Gamma^2/4)$ and $\delta_s = 4g^2_{ens} \Gamma (\omega_s-\omega_{\mathrm{r}})/((\omega_s-\omega_{\mathrm{r}})^2+ \Gamma^2/4)$, and $r=2$ when measuring in reflection, and $1$ when measuring in hanger geometry.

Eq.~\ref{resoResp} shows clearly that the spins create an additional decay channel for the resonator, which depends on the spin frequency. Integrating the losses over the entire spin linewidth probed in magnetic field thus yields:
\begin{equation}
\int_{0}^{+\infty} \frac{\kappa_{\mathrm{s}}}{\omega_{\mathrm{r}}} d\omega_s = \frac{4 \pi g_{ens}^2}{\omega_{\mathrm{r}}}
\end{equation}
\subsubsection{Magnetic field participation ratio}
These spin induced losses can be shown to be proportional to a magnetic field participation ratio, similarly to dielectric losses being proportional to an electrical participation ratio \cite{wangSurfaceParticipationDielectric2015}.
On one hand, we can calculate the magnetic field participation ratio as:
\begin{equation}
    p = \frac{\frac{1}{\mu_0}\int_{V_s} |\delta B_\perp|^2 dV}{\frac{1}{\mu_0}\int_V |\delta B|^2 dV} = \frac{2}{\mu_0\hbar\omega_{\mathrm{r}}} \int_{V_s} |\delta B_\perp|^2  dV
\end{equation}
where $V_s$ is the volume containing the spins of interest (either at the surface of the sapphire, or the surface of the NbTiN), and $V$ is the entire mode volume. Assuming the Zeeman effect predominates in the spin Hamiltonian and that the spin system behaves closely to a spin 1/2,  only the ac-field components orthogonal to the static field $B_0$ will induce a coupling to the spin, we thus only consider $|\delta B_\perp|^2$ for calculating the magnetic field participation ratio.

On the other hand, for a resonator-spin transverse coupling, the spin collective coupling constant can be calculated as :
\begin{equation}
    g_{ens}^2 = \int_{V_s} c |\delta B_\perp|^2 \gamma_e^2 |\langle 1 | S_x/2 | 2 \rangle|^2 dV
\end{equation}
where $c$ is the spin volume concentration, $\gamma_e$ is the spin transition gyromagnetic ratio, and $S_x$ the dimensionless spin operator. We thus find:
\begin{equation}
\int_{0}^{B_0^{\mathrm{max}}} \frac{\kappa_{\mathrm{s}}}{\omega_{\mathrm{r}}} dB_0 = h \mu_0 c \gamma_e |\langle 1 | S_x/2 | 2 \rangle|^2 p,
\end{equation}
giving a linear relation between peak area and participation ratio. To evaluate numerically the participation ratios for all designs, we performed microwave simulations (using HFSS Ansys software) modelling the NbTiN as a perfect conductive layer without any thickness. We extract the contribution of two zones, the free sapphire surface $\Sigma_f$ and the NbTiN surface $\Sigma_{sc}$, to the overall magnetic mode energy $U_B = \frac{\int_V |B|^2 dV}{2\mu_0}$. We use the included field calculator to evaluate $p_\Sigma = t \int_\Sigma |B_{\perp}|^2 d\Sigma / (2\mu_0 U_B)$, where $t$ represents the thickness of the considered surface. We take $t =$\qty{3}{nm}, which is a fair assumption for radicals contribution at the sapphire surface \cite{unNatureDecoherenceQuantum2022}, but a rather arbitrary choice for the NbTiN surface (that nevertheless allows for convenient comparison to the sapphire surface). This modeling also neglects the inhomogeneous current distribution within a superconducting sheet.
\bibliography{NbTiNSapphire_losses}% Produces the bibliography via BibTeX.

\end{document}